\documentclass[prl,showpacs,footinbib,twocolumn,superscriptaddress,final]{revtex4}
\usepackage{graphics,graphicx}

\def\tauphi{$\tau_{\phi}$ } \def\lphi{$l_{\phi}$ }
 
\def\be{\begin{equation}} \def\ee{\end{equation}}

\begin{document}

\title{Experimental Test of the Numerical Renormalization Group Theory for Inelastic Scattering from Magnetic Impurities}

\author{Christopher B\"auerle$^1$, Fran\c{c}ois Mallet$^1$, Felicien Schopfer$^{1,\star}$, Dominique Mailly$^2$, Georg Eska$^3$, and Laurent Saminadayar}

\address{Center for Low Temperature Research, CRTBT-CNRS, BP 166 X, 38042 Grenoble Cedex 9 France, \\
%$^1$Universit\'{e} Joseph Fourier, B.P. 53, 38041 Grenoble Cedex 9, France, \\
$^2$Laboratoire de Photonique et Nanostructures, route de Nozay, 91460 Marcoussis, France, \\
$^3$Physikalisches Institut, Universit\"at Bayreuth,
Universit\"atsstrasse 30, 95440 Bayreuth, Germany \\}

\date{\today}

\begin{abstract}

We present measurements of the phase coherence time \tauphi in quasi
one-dimensional Au/Fe Kondo wires and compare the temperature
dependence of \tauphi with a recent theory of inelastic scattering
from magnetic impurities (Phys. Rev. Lett. 93, 107204 (2004)). A
very good agreement is obtained for temperatures down to 0.2\,$T_K$.
Below the Kondo temperature $T_K$, the inverse of the phase
coherence time varies linearly with temperature over almost one
decade in temperature.

\end{abstract}


\maketitle

The Kondo problem has been fascinating physicist for more than 30
years. Such a large interest is due to the fact that this model,
first introduced to describe transport properties of metals
containing magnetic impurities, is a generic model for the
description of many solid state physics phenomena\cite{hewson}.
Concerning magnetic alloys, the "historical" Kondo solution first
allowed to describe the transport properties at temperatures larger
than the characteristic temperature of the model, the Kondo
temperature $T_{K}$\cite{kondo_64}. The major breakthrough has been
the pioneering work of Wilson who was able to calculate the ground
state of the Kondo problem at all temperatures using his Numerical
Renormalisation Group (NRG)\cite{wilson}. Another important
contribution was due to Nozi\`{e}res who showed that the zero
temperature limit of the Kondo model can be described within
Landau's Fermi liquid theory\cite{nozieres_jltp_74} which made
precise predictions for transport as well as thermodynamic
properties. Since then, many experiments have confirmed the
relevance of this approach.

Recently, it has been suggested that scattering by magnetic
impurities\cite{birge_prl_02,saclay_prb_03} is responsible for the
experimentally observed saturation of the phase coherence time
\tauphi of electrons in metals at low
temperatures\cite{mohanty_prl_97}, renewing the interest in Kondo
physics. Though the Altshuler-Aronov-Khmelnitzky (AAK) theory
describes the temperature dependence of the phase coherence time of
electrons in pure metals\cite{AAK_82}, no exact solution is
available for the phase coherence time in the presence of Kondo
impurities. Only a high temperature expansion, the Nagaoka-Suhl (NS)
expression, was able to describe the experimental data at
temperatures $T\,\gg\,T_{K}$\cite{gruner}. In the opposite limit
($T\,\ll\,T_{K}$), Fermi liquid theory predicts a $T^2$ dependence
of the inelastic scattering rate\cite{nozieres_jltp_74}. This,
however, has never been observed experimentally. Only very recently,
Zar\'{a}nd and coworkers have been able to obtain an exact solution
for the inelastic scattering time in Kondo metals using Wilson's
NRG\cite{zarand_prl_04}. This calculation constitutes a major
breakthrough for the decoherence problem, as it allows to compare
experimental data with theoretical results \emph{for all
temperatures, ranging from well above $T_{K}$ down to zero
temperature}. It is thus of crucial importance to check whether this
theory can describe correctly experimental data, as it could give
new insights into the problem of the low temperature decoherence in
mesoscopic metallic wires.

In this Letter, we compare this recent theory with experimental
results and show that the NRG calculation describes very well the
experimental data at temperatures around and below the Kondo
temperature. In the first part of this Letter we analyze the phase
coherence time of quasi one-dimensional gold wires containing
magnetic iron impurities of ref.\cite{mohanty_prl_00}, denoted as
AuFe1 and ref.\cite{schopfer_prl_03}, denoted as AuFe2, whereas in
the second part we use the NRG calculation to interpret the phase
coherence time as a function of temperature of an extremely pure
gold sample (Au1). In table \ref{Table}, we list the geometrical and
electrical parameters of these samples.

\begin{figure}[h]
\includegraphics[width=6.8cm]{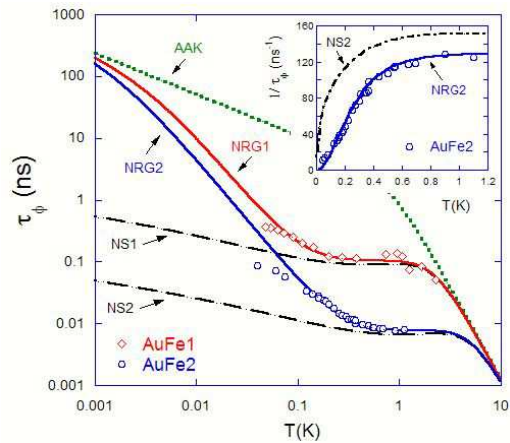}
  \caption{(color online) Phase coherence time as a function of temperature for two Au/Fe Kondo
  wires: $\diamond$ sample AuFe1 of ref.\cite{mohanty_prl_00}, $\circ$ sample AuFe2 (sample B) of ref.\cite{schopfer_prl_03}.
  The green dotted line (denoted AAK) corresponds to the assumption that only electron-electron and electron-phonon interaction contribute to dephasing.
  The black dashed-dotted lines (NS1 and NS2) take in addition account for the magnetic scattering using the NS expression, whereas for the red (NRG1) and
  blue (NRG2) solid lines, the NRG calculation has been employed for the contribution of the magnetic impurities. The inset shows 1/\tauphi versus T on a linear
  scale for sample AuFe2 in order to emphasize the \textit{linear} regime.}
    \label{NRG}
 \end{figure}

\begin{table}[h]
\squeezetable
\begin{tabular}{c c c c c c c c c}
{Sample}&{$w$}&{$t$}&{$l$}&{$R$}&{$D$}&{$c_{imp}$}\\
{}&{(nm)}&{(nm)}&{($\mu$m)}&{($\Omega$)}&
{(cm$^{2}$/s)}&{$(ppm)$}\\
\hline
\vspace{-2mm} \\
{AuFe1}&{$180$}&{$40$}&{$155$}&{$393$}&{$200$}&{$3.3$}\\
{AuFe2}&{$150$}&{$45$}&{$450$}&{$4662$}&{$56$}&{$45$}\\
{Au1}&{$120$}&{$50$}&{$450$}&{$1218$}&{$241$}&{$<0.015$}
\end{tabular}
\caption{Sample characteristics: $w, t, l, R,$ correspond to the
width, thickness, length and electrical resistance, respectively.
$D$ is the diffusion coefficient and $c_{imp}$ is the impurity
concentration extracted from the NRG fits.} \label{Table}
\end{table}
The phase coherence time of sample AuFe1 and AuFe2, as extracted
from standard weak localization measurements is shown in figure
\ref{NRG}. Both samples display a distinct plateau at a temperature
above $T\,\geq\,0.3\,K$. This plateau is caused by Kondo spin flip
scattering due to the presence of iron impurities which leads to
dephasing. Decreasing the temperature, the magnetic impurity spin is
screened and as a result, the phase coherence time increases again.

The temperature dependence of the measured phase coherence time
\tauphi in the presence of magnetic impurities can be described in
the following way
\begin{equation}
\frac{1}{\tau_\phi} = \frac{1}{\tau_{e-e}} + \frac{1}{\tau_{e-ph}}+
\frac{2}{\tau_{mag}}
\end{equation}
where $1/\tau_{e-e}\,=\,a_{theo}\,T^{2/3}\,=\,[\frac{e^2 R \sqrt{D}
k_B}{2^{3/2} \hbar^2 L}]^{2/3} \, T^{2/3}$ corresponds to the
electron-electron interaction term\cite{gilles_book},
$1/\tau_{e-ph}\,=\,b\,T^{3}$ to the electron-phonon interaction,
while $1/\tau_{mag}$ correspond to the contribution due to magnetic
impurities. To account for the scattering off magnetic impurities,
we use on one hand the common Nagaoka-Suhl
expression\cite{haesendonck_prl_87,gruner}
\begin{equation}
\frac{1}{\tau_{mag}} = \frac{1}{\tau_{NS}} =
\frac{c_{imp}{[ppm]}\quad \quad \pi^2
S(S+1)}{0.6\,{[ns]}\,\,\,(\pi^2 S(S+1) + ln^2(T/T_K))}
\end{equation}
where $S$ is the impurity spin, $c_{imp}$ the impurity concentration
expressed in parts-per-million(ppm) and the prefactor of 0.6\,[ns]
has been calculated taking the electron density of
gold\cite{saclay_prb_03}. On the other hand we use the inelastic
scattering rate calculated by NRG\cite{zarand_prl_04}
\begin{equation}
\frac{1}{\tau_{mag}} = \frac{1}{\tau_{inel(NRG)}} = A \frac{\sigma
(w)_{inel}}{\sigma_0}*c_{imp}
\end{equation}
where $\sigma (w)_{inel}$ is the inelastic scattering cross section
at finite energy $w$, $\sigma_0 = 4\pi/k_F^2$ the elastic scattering
cross section at zero temperature and $A$ a numerical constant in
units of [s$^{-1}$] to express the impurity concentration $c_{imp}$
in ppm.

The green dotted line in figure\,\ref{NRG}, denoted as AAK,
corresponds to the assumption that only electron-electron and
electron-phonon interaction contribute to the electron dephasing and
that there is no other mechanism for decoherence at low
temperatures. For the simulation we have used
$a_{theo}\,=\,0.41\,ns^{-1}K^{-2/3}$ and $b\,=\,0.
8\,ns^{-1}K^{-3}$. Keeping these values fixed, and adding the
contribution due to magnetic impurities using the NS expression for
S=1/2, we obtain the black dashed-dotted lines for sample AuFe1 and
AuFe2, denoted as NS1 and NS2. The NS expression describes
relatively well the observed temperature dependence of \tauphi at
temperatures above $T_K$ but fails to describe the desaturation of
\tauphi due to the screening of the magnetic impurities at
temperatures $\leq T_K$. This is not at all surprising since the NS
expression results from a perturbative expansion in $(T/T_K)$ and
breaks down for $T\rightarrow T_K$. Instead, taking the inelastic
scattering rate obtained by NRG we obtain a very good agreement with
the experimental data in basically the entire temperature range as
shown by the red (NRG1) and blue (NRG2) solid lines for sample AuFe1
and AuFe2, respectively. For the fitting procedure, we have adjusted
the magnetic impurity concentration such that the NS expression and
the NRG calculation coincide at high temperatures ($T>10\,T_K$). One
clearly sees that the NS expression deviates already from the NRG
data at relatively high temperatures ($T \sim 5\,K$) as already
pointed out in ref.\cite{zarand_prl_04} and fitting of experimental
data with this expression at temperatures $T < 10\,T_K$ should
therefore be avoided. From the NRG fitting procedure we obtain an
impurity concentration of approximately 3.3\,ppm and $T_K =
0.4\,K\pm 0.05\,K$ for sample AuFe1 and 45\,ppm and $T_K = 0.9\,K
\pm 0.05\,K $ for sample AuFe2\cite{comment_T_K}, in good agreement
with $T_K$ values for Au/Fe found in the literature\cite{rizutto}.
It is worth mentioning that the measurement of the phase coherence
time is a very precise method to determine the Kondo temperature as
well as the magnetic impurity concentration. In resistance
measurements, the determination of the Kondo temperature is not
straightforward since the entire temperature range down to the
unitary limit ($T\ll T_K)$ is necessary to extract the Kondo
temperature with a satisfactory precision. The NRG fitting procedure
of \tauphi on the contrary allows to extract the Kondo temperature
with high precision if temperatures only slightly lower than $T_K$
are attained.

At the lowest temperatures we observe deviations from the NRG
theory. A probable explanation for the deviation is fact that the
spin of Fe in Au is not exactly 1/2 and the impurity spin might not
be completely screened. This eventually leads to interactions of
magnetic impurities and a saturation of
\tauphi\cite{schopfer_prl_03}. Another possibility is the presence
of another magnetic impurity with a much lower Kondo temperature.
This latter issue will be discussed in the last part of this
article.

The temperature evolution of \tauphi below $T_K$ deserves several
comments: after a slow increase of \tauphi the temperature
dependence is almost \textit{linear} in temperature over almost one
decade in temperature, as emphasized in the inset of figure
\ref{NRG}. This relatively weak temperature dependence explains the
fact why the pioneering
experiments\cite{bergmann_prl_87,haesendonck_prl_87} have not
succeeded to observe the Fermi liquid regime. Comparing the
experimental results with the NRG calculation, we clearly see that
the Fermi liquid regime can only be reached for temperatures
typically below $0.01 \,T_K$. Moreover, the AAK behaviour is only
recovered at extremely low temperatures ($T<0.001\,K$).

It is noteworthy, that the calculated quantity by NRG is
$\sigma_{inel}$ and not the phase coherence time \tauphi measured in
a weak localization experiment. In fact, $\sigma_{inel}$ has been
calculated in the limit of zero temperature and finite energy
$\sigma_{inel}(w,T=0)$, whereas in a transport experiment one
measures $\sigma_{inel}(w=0,T)$. The fact that the numerical results
describe this well the experimental data let us conclude that these
two quantities are not very different, at least for
$k_BT\ll\epsilon_F$, $\epsilon_F$ being the Fermi energy.

\begin{figure}[h]
\includegraphics[width=7cm]{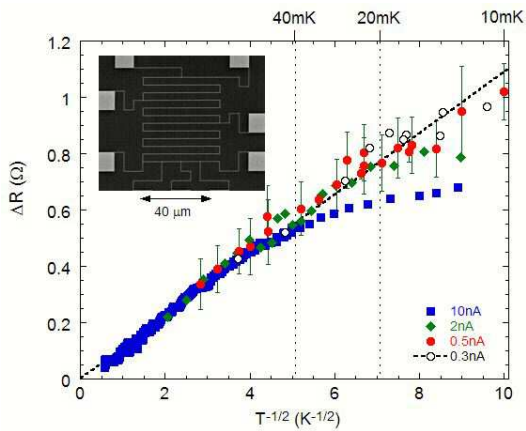}
  \caption{(color online) Resistance variation of sample Au1 plotted as a function of $1/\sqrt{T}$ for different bias currents.
  The dotted line corresponds to the theoretical expectation for the resistance correction.
  The inset shows a SEM photograph of the gold wire.}
\label{R_T}
\end{figure}

Having now a theory which satisfactorily describes the temperature
dependence of \tauphi in the presence of magnetic impurities, let us
reexamine the temperature dependence of \tauphi in extremely pure
gold wires. In a recent article\cite{saclay_prb_03}, the deviation
of \tauphi from the AAK prediction at very low temperatures has been
assigned to the presence of an extremely small amount of magnetic
impurities (typically on the order of 0.01ppm). For this purpose we
have fabricated an extremely pure gold wire (Au1) as shown in the
inset of figure \ref{R_T}. The fabrication procedure is essentially
the same as for the Au/Fe wires with the only difference that the
wire has been evaporated in an evaporator which is exclusively used
for the evaporation of extremely pure gold. The gold of purity 5N5
has been evaporated directly on a silicon wafer $without$ a sticking
layer. In addition, special care has been taken for the sample
design such that there is no influence on the phase coherence due to
the two dimensional contact pads (see inset of figure 2). For this
wire the phase coherence length at the lowest temperatures is more
than $20\mu m$. To our knowledge, this is the largest coherence
length ever obtained in a metallic wire and confirms the high purity
of the sample.

To determine the effective electron temperature of this sample, we
have measured the Altshuler-Aronov correction to the resistivity at
very low temperatures. A magnetic field of 40\,mT has been applied
in order to suppress weak localization correction to the
resistivity. In figure \ref{R_T} we plot the resistance correction
as a function of $1/\sqrt{T}$. For measuring currents below 0.7\,nA,
the sample is in thermal equilibrium ($eV<kT$) in the entire
temperature range and the resistance correction follows the expected
$1/\sqrt{T}$ temperature dependence down to 10\,mK. This clearly
shows that the electrons of a mesoscopic sample can be cooled to
such low temperatures. Fitting the temperature dependence of the
resistance correction to $\Delta R(T)=\alpha_{exp}/\sqrt{T}$ (dotted
line in figure\,\ref{R_T}), we determine $\alpha_{exp}$ and compare
it to the predicted value \cite{gilles_book} of $\Delta
R(T)=2R^2/R_K L_T/L=\alpha_{theo}/\sqrt{T}$, where $L_T=\sqrt{\hbar
D/k_B T}$ is the thermal length and $R_K=h/e^2$. We obtain a value
$\alpha_{exp}\,=\, 0.11\,\Omega /K^{1/2} $ which is in very good
agreement with the theoretical value of
$\alpha_{theo}\,=\,0.109\,\Omega /K^{1/2}$.

\begin{figure}[h]
\includegraphics[width=7cm]{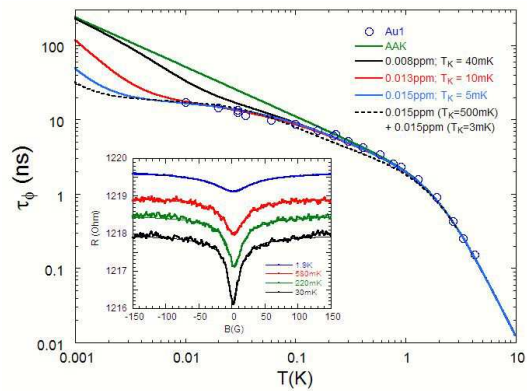}
  \caption{(color online) Phase coherence time as a function of temperature for sample Au1 ($\circ$).
 The solid green line corresponds to the AAK prediction,
 the black (a), red (b) and blue (c) solid lines correspond to the NRG calculation assuming $T_K=40\,$mK,
 $T_K=10\,$mK, and $T_K=5\,$mK, respectively.
 The inset shows typical magneto-resistance curves at different
temperatures.}
    \label{NRG_webb}
\end{figure}

The phase coherence time \tauphi is then measured via standard weak
localization (see inset of figure \ref{NRG_webb}) and the phase
coherence length \lphi is extracted via the Hikami-Larkin-Nagaoka
formula\cite{hikami_80}. From the relation $\tau_\phi = \l_\phi^2/D$
we then calculate the phase coherence time as displayed in figure
\ref{NRG_webb}. We fit the experimental data with the AAK expression
such that an almost perfect agreement is obtained at high
temperatures $(T>100\,mK)$, as shown by the green solid line. The
prefactor  we extract from this fit $a_{fit}=0.42 \,ns^{-1}K^{-2/3}$
is in very good agreement with the theoretical prediction of
$a_{theo}\,=\,0.41\,ns^{-1}K^{-2/3}$. At temperatures below 100\,mK
our data deviate substantially from the AAK prediction. To see
whether these deviations can be explained by the presence of a very
small amount of magnetic impurities, we simulate the temperature
dependence of \tauphi for the presence of a small amount of magnetic
impurities using the NRG calculations. The black (a), red (b) and
blue (c) solid lines correspond to a simulation assuming
$T_K$\,=\,40\,mK, $T_K$\,=\,10\,mK, and $T_K$\,=\,5\,mK, with an
impurity concentration of $c_{imp}$\,=\,0.008\,ppm,
$c_{imp}$\,=\,0.013\,ppm, and $c_{imp}$\,=\,0.015\,ppm,
respectively. It is clear from our simulations that only magnetic
impurities with a Kondo temperature $T_K \leq$\,10mK and with a
concentration smaller than 0.015\,ppm describe satisfactorily the
experimental data. A possible magnetic impurity with a Kondo
temperature in this temperature range is Mn
($T_K\,\simeq\,$3mK)\cite{eska}.

For the sake of objectiveness let us point out, however, the
following: assuming magnetic impurities with a Kondo temperature
below the measuring temperature leads to an almost temperature
independent scattering rate for $T\geq T_K$. Any experimentally
observed saturation of \tauphi can therefore always be assigned to
magnetic impurities with an unmeasurably low Kondo temperature. One
could also argue, that it is curious that for the case of gold
wires, the observed temperature dependence of \tauphi can only be
described satisfactorily by assuming the presence of one specific
magnetic impurity with a Kondo temperature below the measuring
temperature $T\leq $\,10mK, whereas it is known that the dominant
magnetic impurity in gold is iron. If we assume an additional iron
concentration (0.015\,ppm; $T_K$\,=\,500\,mK) of the same order as
for instance Mn (0.015ppm; $T_K$\,=\,3\,mK), the temperature
dependence of \tauphi does not satisfactorily describe the
experimental data as displayed by the dotted black line (d). A
possible explanation of these facts, might be the presence of a
distribution of Kondo temperatures, which could be significant for
very diluted Kondo impurities \cite{kettemann}. Such a distribution
of Kondo temperatures has already be seen in point contact
experiments\cite{yanson}. This, however, can only be verified with
phase coherent measurements at high magnetic
fields\cite{mohanty_prl_03}. We also note that our present results
do not allow to rule out the predictions of ref. \cite{zaikin}.

In conclusion we have shown that the NRG calculation of the
inelastic scattering rate describe very well the experimentally
observed temperature dependence of \tauphi caused by the presence of
magnetic impurities. Below $T_K$ the inverse of the phase coherence
time varies basically linearly with temperature over almost one
decade in temperature.  The $T^2$ temperature dependence predicted
by the Fermi liquid theory, on the other hand, can only be reached
for temperatures smaller than $0.01\,T_K$ and remains an
experimental challenge.
\\note added in proof: an exact calculation of \tauphi in presence of
disorder supports the above findings \cite{rosch}.
\section{Acknowledgelents}
We acknowledge helpful discussions with P. Simon, S. Kettemann, G.
Zar\'and, A. Rosch, G. Montambaux, C. Texier, H. Bouchiat and L.P.
L\'evy. In addition, we would like to thank G. Zar\'and for
providing us with the numerical data of the NRG calculation of
$\sigma_{inel}$. We are also indebted to R.A. Webb for the gold
evaporation of sample Au1. This work has been supported by the
French Ministry of Science, grants \# 02 2 0222 and \# NN/02 2 0112,
the European Comission FP6 NMP-3 project 505457-1 \textquotedblleft
Ultra-1D\textquotedblright, and by the IPMC Grenoble. C.B. and G.E.
acknowledge financial support from PROCOPE.


\begin{thebibliography}{99}
\bibitem{comment_T_K} The fact that the extracted Kondo
temperatures are different for the two samples is somewhat
surprising. However, these results are consistent with the maxima
observed previously in the spin flip rate\cite{schopfer_prl_03,
bergmann_prl_87}.
\end{thebibliography}
\end{document}